# Generative Reversible Data Hiding by Image to Image Translation via GANs

Zhuo Zhang[1,2], Guangyuan Fu[1], Fuqiang Di*[2], Changlong Li[3], Jia Liu[2]

1 Xi'an Research Institute of High Technology, Xi'an, 710086, China

2 Key Lab of Networks and Information Security of PAP, Xi'an, 710086, China

3 The general staff of PAP, Beijing, 100000, China

Correspondence should be addressed to Fuqiang Di; 18710752607@163.com

**Abstract:** The traditional reversible data hiding technique is based on cover image modification which inevitably leaves some traces of rewriting that can be more easily analyzed and attacked by the warder. Inspired by the cover synthesis steganography based generative adversarial networks, in this paper, a novel generative reversible data hiding scheme (GRDH) by image translation is proposed. First, an image generator is used to obtain a realistic image, which is used as an input to the image-to-image translation model with CycleGAN. After image translation, a stego image with different semantic information will be obtained. The secret message and the original input image can be recovered separately by a well-trained message extractor and the inverse transform of the image translation. Experimental results have verified the effectiveness of the scheme.

**Keywords:** generative reversible data hiding, image to image translate, generative adversarial networks

## 1 Introduction

Information hiding [1-5], also called data hiding, is an important information security technique widely used in secret transmission [6], digital copyright protection [7], and other scenarios. If we classify data hiding by the reversibility of cover image, data hiding can generally be divided into two types: irreversible data hiding (IDH) [8-9] and reversible data hiding (RDH) [10-12]. The traditional data hiding methods can be classified into the former type while the latter type is a special technique which is mainly applied to medical, judicial, and military fields.

With the emergence and development of artificial intelligence [13-16] and other new techniques, the IDH methods using deep learning models have achieved a series of breakthrough in methods and performance, and become the trend of development in this field [17-20]. Among these new methods, secret data can be hidden and extracted well without any modification in the original cover image, and cannot be detected by the warder (steganalysis algorithm). Comparatively, RDH with deep learning has received less attention. As far as our best knowledge, there is currently no RDH method can hide data without modification. At present, RDH methods can be divided into two types: RDH in unencrypted images [21-23] and RDH in encrypted images [24-25]. Among these RDH methods, data hiding is based on cover image modification, which is more and more easily to be detected by increasingly advanced machine learning detection tools.

In [18], a new image steganography method via deep convolutional generative adversarial networks (DCGAN) is proposed. In this method, a mapping from the secret data to random noise is designed. With this mapping, a corresponding relationship between secret data and the stego image generated by DCGAN model is obtained, and an extractor used to extract secret data is trained. This method has a strong ability to resist state-of-the-art detection tools, and it has provided great inspiration for RDH method without modification.

Cycle-Consistent Generative Adversarial Networks (CycleGAN) [26] is a newly proposed image to image translate model, which learns to automatically translate an image from a source domain into a target domain in the absence of paired examples. In this generative adversarial network (GAN) model, there are two generators and two discriminators. Cycle-consistency loss is defined to train CycleGAN model. Using CycleGAN, one type of picture can be transformed into another, and this transformation is reversible. Obviously, this kind of technique can be applied to RDH field.

In [27], a framework for RDH in encrypted images based on reversible image transformation is proposed. At first, a cover image is transformed into another target image by image transformation. Then, secret data is embedded into the transformed image, which regarded as encrypted image. There have been many similar methods [28-30]. In this type of method, the image transformation is regarded as a special type of image encryption. And the embedding method belongs to the traditional method and is essentially relied on pixel modification. However, the generative model uses neural networks to learn the data distribution rule of samples, and the generated image has strong randomness, which enhances the security of the data hiding algorithms. This advantage is far superior to traditional methods.

In this paper, the generative reversible data hiding (GRDH) method based on

GANs model is proposed. Learning the secret data mapping method in [18] and the image recovery method in CycleGAN, a new GRDH framework is proposed. In this framework, a cover image is generated by a noise vector, which is transformed by the secret data. Then, the cover image is transformed into a marked image by CycleGAN model. Similar to the frame in [27], the transformed image can be regarded as a special encrypted image. In addition, a new extractor is trained to extract the secret data, which make the data hiding framework reversible. Experimental results have proved the feasibility of the proposed GRDH method.

**2 DCGAN and CycleGAN**

DCGAN [31] is an upgraded version of GAN. In DCGAN model, convolution neural network is introduced to design generator and discriminator. To improve the quality of generated samples and the speed of convergence process, some changes to the structure of original convolution neural network have been made. With the powerful feature extraction ability of convolution neural network, the learning effect of the generative adversarial network has been significantly improved. Illustration of DCGAN model has been shown in Fig. 1. The fake image is generated by random noise and a generator. The discriminator is designed in order to judge whether the generated image is real image or fake image. The goal of generator is to generate real images to deceive discriminator. The goal of discriminator is to separate the fake image from the real one as far as possible. In this way, generator and discriminator constitute a dynamic game process.

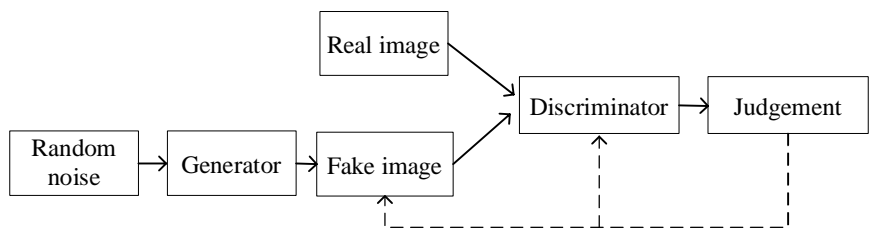

**Fig. 1** Illustration of DCGAN model

CycleGAN [26] is an essentially ring network which is made up of two symmetrical GAN models. Illustration of CycleGAN model has been shown in Fig. 2. On the one hand, two GAN models share two generators. On the other hand, each GAN model contains a discriminator. Thus, there are two generators and two discriminators in a CycleGAN model. Fig. 3 shows the illustration of a one-way GAN model. Real images in domain A can be transformed to fake images in domain B based on a discriminator, and then transformed to recovered image of domain A by another generator.

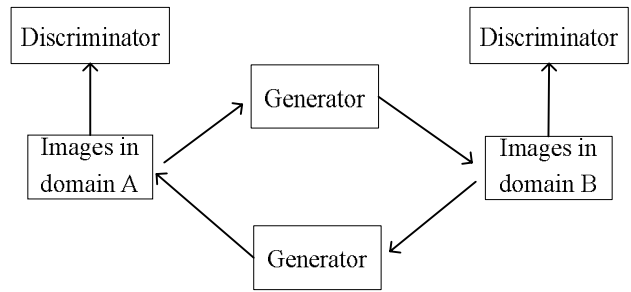

**Fig. 2** Illustration of CycleGAN model

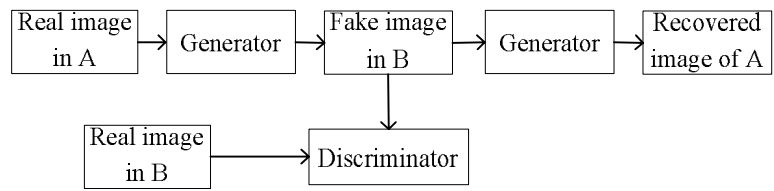

**Fig. 3** Illustration of a one-way GAN in CycleGAN

## 3 Generative Reversible Data Hiding

The illustration of GRDH has been shown in Fig. 4.

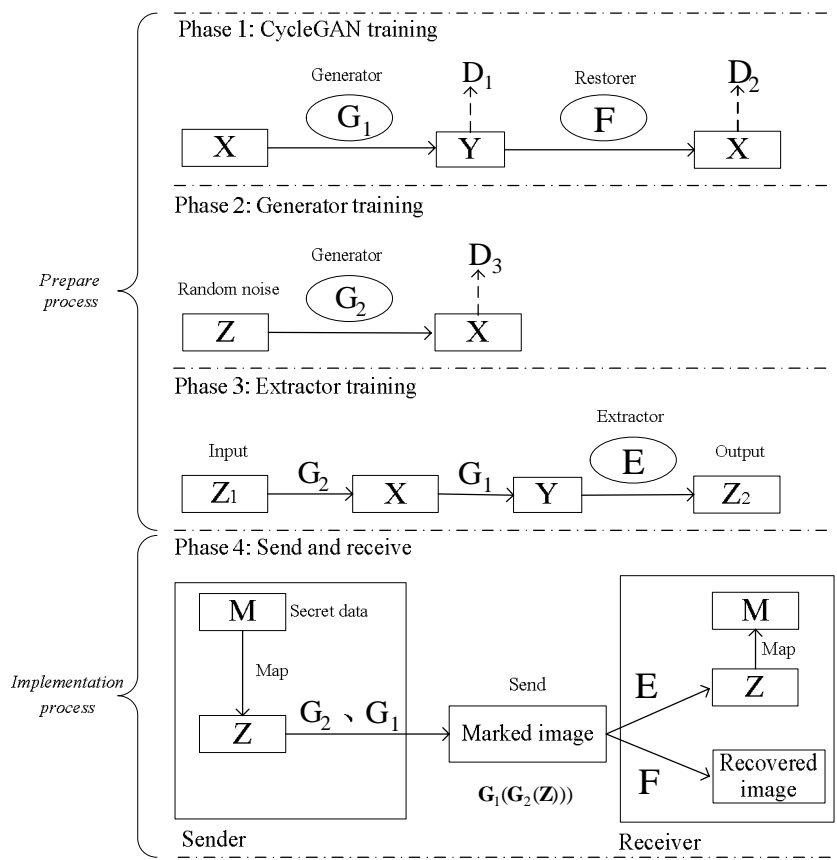

**Fig. 4** Illustration of the proposed GRDH method

There are two processes in GRDH: prepare process and implementation process, consisting of 5 phase:

Phase 1, CycleGAN training. A generator $G_1$ and a restorer $F$ are generated by CycleGAN method. With two discriminators $D_1$ and $D_2$, two image mapping goals are achieved: $X \to Y$ and $Y \to X$, where $X$ and $Y$ are image collections.

Phase 2, Generator training. A generator $G_2$ is obtained by a GAN method (e.g. DCGAN or BEGAN) with the help of discriminator $D_3$.

Phase 3, Extractor training. This phase based on the two discriminators $G_1$ and $G_2$ obtained earlier, we can achieve the transformation from random noise to image collection $Y$. Then, we train a new extractor $E$ based on GAN technique and ensure that the generated output $Z_2$ is same with the input $Z_1$ as closely as possible.

Phase 4, Send and receive. Before data hiding, the sender sends extractor $E$ and restorer $F$ to the receiver. Both sides learn a mapping from secret data $M$ to noise $Z$. Corresponding to the traditional RDH methods, the image generated by $G_1$ and $G_2$ can be regarded as cover image and marked image. Then, the sender sends the marked image $G_1(G_2(Z))$ to the receiver. At the receiver side, recover image can be obtained and the embedded data can be extracted. We will go into detail about the various phases in the following.

*3.1 Prepare process*

The purpose of the prepare process is to train related models and prepare for data hiding, including three phases: CycleGAN training, generator training, and extractor training. The detailed steps can be described as follows:

Phase 1: CycleGAN training

Our goal of this phase is to generate a marked image generator $G_1$ and a recovered image restorer $F$. Without loss of generality, we choose the original CycleGAN model to train. Assume $X$ and $Y$ denote two image collections, which corresponding to cover image and marked image respectively. Firstly, build two image databases: a cover image database ($CDB$) and a marked image database ($MDB$). Each database contains one type of images. For example, $CDB$ contains images of normal horses and $MDB$ contains images of zebras. Then, the cover image in the following phases will be an image of normal horse and the marked image sending to the receiver can be an image of zebra. In Phase 1, the training process is based on the original CycleGAN model. We apply the two adversarial losses: $L_{GAN}(G_1, D_2, X, Y)$ and $L_{GAN}(F, D_1, Y, X)$ for mapping $X \to Y$ and $Y \to X$ defined in [26], and the full objective function can be described as follows.

$$L(G_1, F, D_1, D_2) = L_{GAN}(G_1, D_2, X, Y) + L_{GAN}(F, D_1, Y, X) + \lambda L_{cyc}(G_1, F) \quad (1)$$

where $L_{cyc}(G_1, F)$ denotes the cycle consistency loss.

Phase 2: Generator training

Our goal of this phase is to generate a cover image generator $G_2$. Without loss of generality, we can choose the original DCGAN model to train. According to the principle of DCGAN model, a generator $G_2$ can be generated by the cover image database $CDB$ and the discriminator $D_3$. Then, the mapping from random noise $z$ to image collection $X$ can be learned. The structures of DCGAN model are introduced in [31]. Both Generator $G_2$ and discriminator $D_3$ are CNN structures. Denote $x$ and $P_{data}$ as the real image and its distribution from cover image database $CDB$, then the objective function to be optimized is as follows.

$$\min\max V(G_2, D_3) = E_{x \sim P_{data}(x)}[\log D_3(x)] + E_{Z \sim P_z(z)}[\log(1 - D_3(G_2(z)))] \quad (2)$$

Other unsupervised GAN models (e.g. BEGAN) can also be used for generator training.

Phase 3: Extractor training

After Phase 1 and Phase 2, two generators $G_1$ and $G_2$ have been trained. Based on these two generators and input noise $z_1$, the marked image can be generated by $G_1(G_2(z_1))$. Our goal of Phase 3 is to train an extractor $E$ for secret data. We draw on Hu's method [18]. The construction method of $E$ is similar to that of the discriminator in DCGAN model, which has four convolutional layers and a fully connected layer. A leak-Relu activation function and batch normalization are used in each layer. Different from conventional CNN models, there is no pooling layer or dropout operation in the extractor. Illustration of the extractor in GRDH method has been shown in Fig. 5.

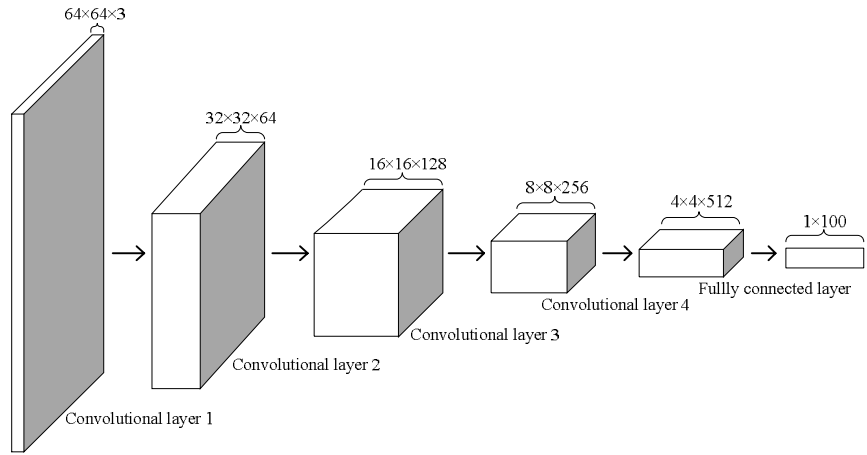

**Fig. 5** Illustration of the extractor in GRDH method

If the input of $E$ is the marked image with the size 64×64×3, the output after the first layer is 32×32×64. In the following layers, the image dimensions are halved, and the number of channels is doubled from the previous layer. The final output is a noise vector of 100 dimensions. Each noise value in the vector is between -1 and 1. The defined loss function for the extractor training can be described as follows.

$$L(E) = \sum_{i=1}^{n}(z_1 - E(G_1(G_2(z_1))))^2 \tag{3}$$

We use this loss function to train the extractor as much as possible so that its output is as close as possible to the input noise $z_1$.

*3.2 Implementation process*

After the prepare process, two generators $G_1$ and $G_2$, an extractor $E$, and a restorer $F$ are obtained. As shown as in Fig. 4, the sender holds $G_1$ and $G_2$ and sends $E$ and $F$ in advance by a secure channel. The above process is similar to that of key distribution in public key cryptography.

According to the steps of prepare process, the noise vector $Z$ is transformed into an image $G_2(Z)$ by $G_2$ at first, and then transformed into another image $G_1(G_2(Z))$ by $G_1$. From the view of RDH technique, the first image $G_2(Z)$ can be seen as the cover image, and the second image $G_1(G_2(Z))$ can be seen as the marked image. In implementation process, the only thing that the image owner needs to do is sending the marked image $G_1(G_2(Z))$ to the receiver. At the receiving end, the receiver can recover the cover image by the restorer $F$, and extract the noise vector by the extractor $E$. From the view of RDH in encrypted images, the above process at the receiving end belongs to separable scheme. It means that the receiver can not only recover the image before data extracting, but also recover the image after data extracting. Beyond that, the mapping method proposed in [18] is used to realize the mapping from the secret binary bits to noise vector. The mapping method can be described as follows.

At first, divide the secret binary bits into several groups. Each group contains $k$ bits. For example, divide {110101100} into three groups {110}, {101} and {100} when $k=3$. Then, map each group to a random noise $r$ with a given interval according to the following equation.

$$r = random(\frac{m}{2^{k-1}} - 1 + \delta, \frac{m+1}{2^{k-1}} - 1 - \delta) \tag{4}$$

Where *m* denotes the decimal value of the group to be mapped, and $\delta$ denotes the gap between the divided intervals. For example, $k=3$, $\delta=0.001$. We map every three secret bits into a random noise with the value between -1 and 1. The mapping from the group to the interval can be shown in Table 1. At last, package all the mapped noise into a vector. The above mapping method allows a deviation tolerance in data extracting, and ensures the extraction accuracy of the secret data during the implementation process. This mapping method will be shared by both the sender and the receiver. The sender maps secret data to noise vector and the receiver maps noise vector to secret data.

**Table 1** The mapping from the group to the interval

| The group | The interval |
| --- | --- |
| 000 | (-0.999,-0.751) |
| 001 | (-0.749,-0.501) |
| 010 | (-0.499,-0.251) |
| 011 | (-0.249,-0.001) |
| 100 | (0.001,0.249) |
| 101 | (0.251,0.499) |
| 110 | (0.501,0.749) |
| 111 | (0.751,0.999) |

**4 Experimental results**

In this section, a group of experiments is conducted to verify the effectiveness of the GRDH method proposed in this paper. These experiments consist of two parts. First, we train the GANs models and the extractor for GRDH preparation. Then, we use these trained models for GRDH to verify the feasibility of the method. In all the experiments, we generate random bits (using random.randint function in NumPy) as secret information. All images in the datasets were resized to $64\times64$ in advance for model training. All experimental results are obtained by Lenovo graphic workstation Think Station P500 with an NVIDIA GeForce GTX 1080Ti GPU and 128 GB of memory.

*4.1 Experimental for GRDH preparation*

4.1.1 CyclyGAN model training

We select the image database Horse and zebra in [32] as training samples. In CycleGAN training stage, we set the initial learning rate to 0.0002 and the least batch size to 100. Besides, the stochastic gradient descent (SGD) [33] is selected as the optimization algorithm in the model training. Some visual experimental results of CycleGAN training have been shown in Fig. 6.

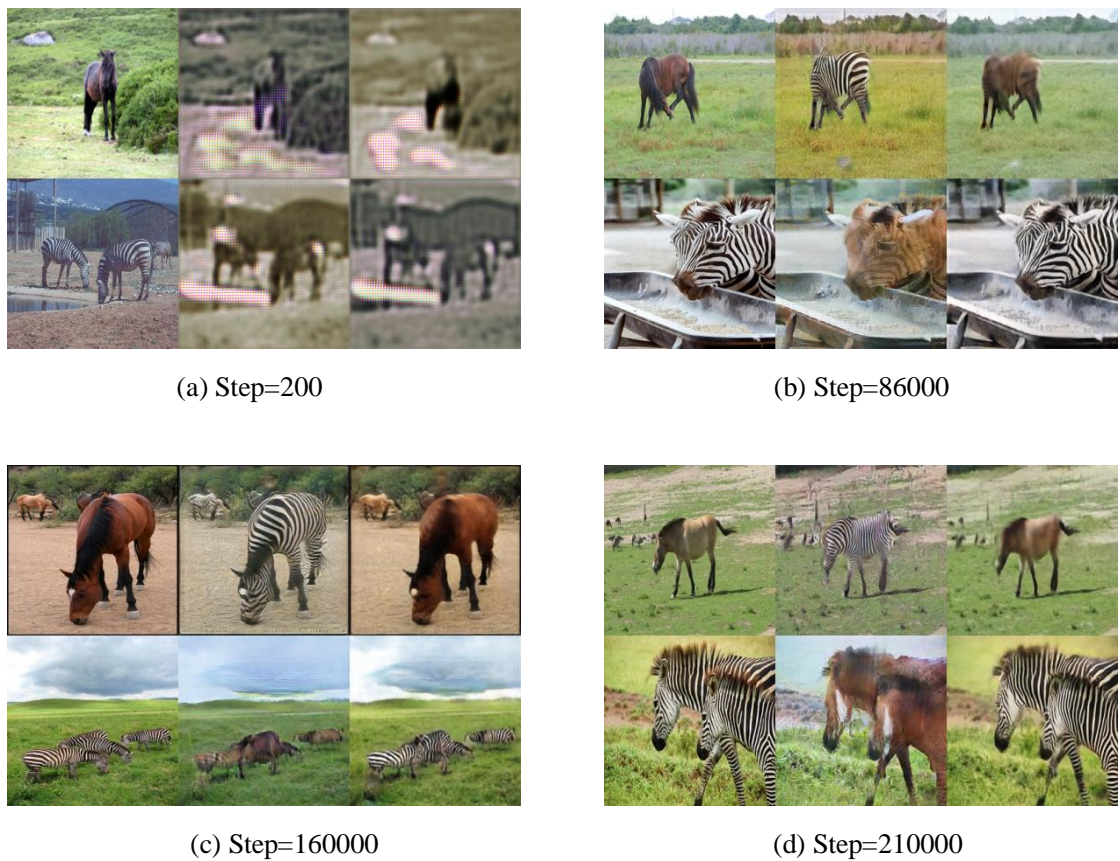

(a) Step=200     (b) Step=86000

(c) Step=160000     (d) Step=210000

**Fig. 6** Partial experimental results of CycleGAN training

The first line stands for the mapping from X domain (horse) into Y domain image (zebra). There images from left to right are respectively the original image, the transformed image and the reconstructed image. The second line stands for the mapping from Y domain image (zebra) into X domain (horse). From Fig. 6, When the model was trained to 210000 steps, the visual results of image transformation and image recovery can be acceptable in some special scenes when the demand for reversibility is not high. The visual results are related with the chosen image database and sample size.

We also use the man2woman image set [34] to train the model. The batch size is set to 100; the random number seed is set to 1234; the initial learning rate initial value is set to 0.002; the rate remains the same in the first 10,000 steps, then decays every 10,000 steps until it decays to zero (one step represents a batch of image training). Assuming that the Man image set is X and the Woman image set is Y, the adjustment parameters are set to 10.0. in the training in both directions X->Y and Y->X. The first moment parameter of the gradient descent optimizer is set to 0.5, and the number of filters of the first convolution layer is set to 64. Fig.7 shows the image quality using the CycleGAN model after different training steps. The CycleGAN model is trained during 600,000 steps and takes about 52 hours, that is, about 11,500 steps per hour. It can be seen from the figure that the quality of the gender-converted image after the number of trainings is more than 100,000 has reached an acceptable level. As the number of trainings continues to increase, the image quality of the CycleGAN model for gender conversion and image restoration is gradually improved.

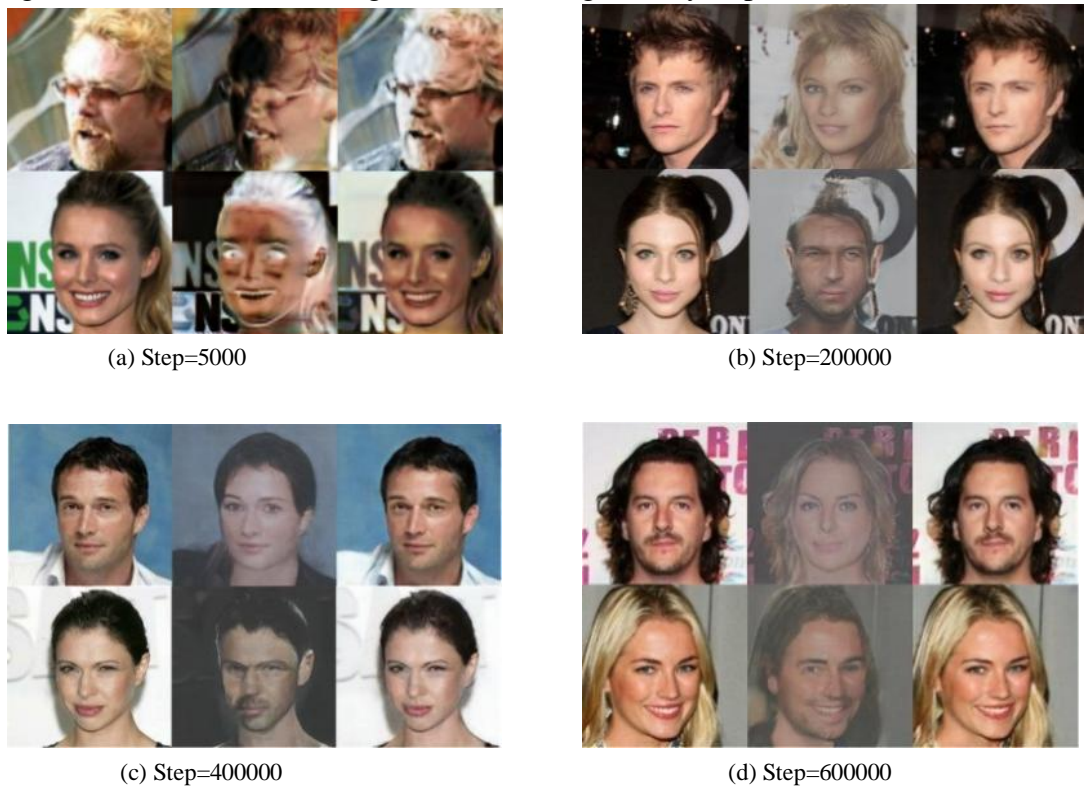

(a) Step=5000         (b) Step=200000

(c) Step=400000       (d) Step=600000

**Fig. 7** CycleGAN model training results

4.1.2 Generator training

Due to the high image quality of BEGAN's production, we directly use BEGAN model for generator training. We use celebA image library [35]; batch size is set to 16; initial base learning rate is set to 0.001, and the learning rate is gradually attenuated. Further, the initial value of the parameter k0=0, and the initial value of the parameter is set to γ=2. The image quality of the image generated by the BEGAN model after different training steps is shown in Fig. 8. It can be seen from the figure that the quality of the generated image gradually increases with the increase of the number of training steps, and the number of steps is gradually increased to a more realistic level with the natural image.

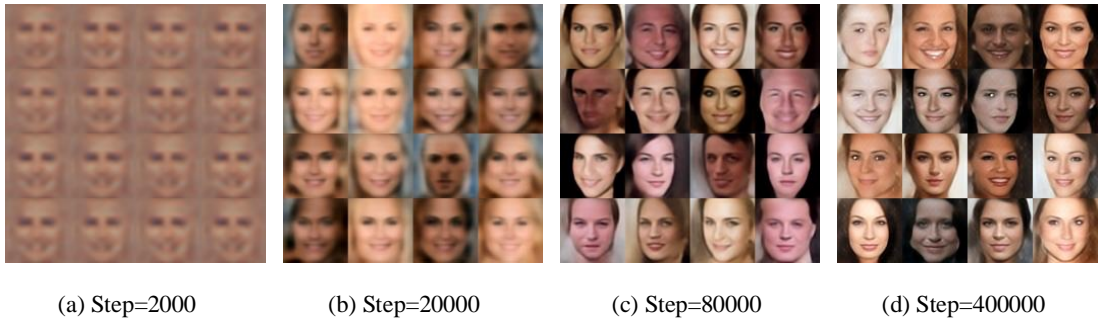

(a) Step=2000    (b) Step=20000    (c) Step=80000    (d) Step=400000

**Fig. 8** BEGAN model training results

It is worth noting that one of the important factors affecting the quality of the generated image is the type of GAN model selected. The BEGAN model is relatively simple, the training time is low, and the image quality is relatively acceptable. Compared with the BEGAN model, although the training of the StyleGAN model takes a lot of time, the generated image is more realistic and closer to the natural image. Fig.9 shows a partial generation of the StyleGAN model based on the FFHQ image library. It can be seen that it is very close to the natural image, but even if the GPU of the NVIDIA Tesla V100 model needs to be trained for about five weeks.

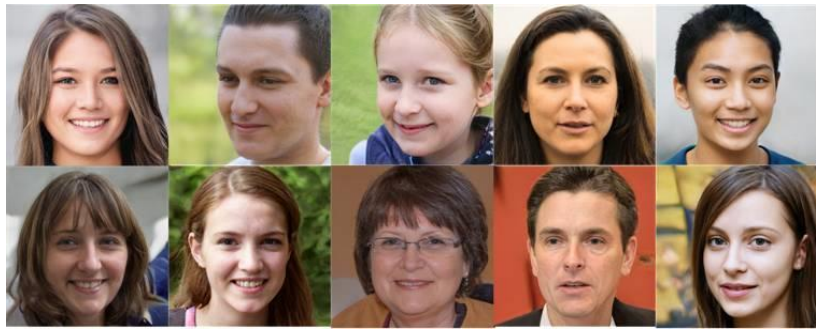

**Fig. 9** StyleGAN model training results

4.1.3 Extractor training

First, we generate 10,000 marked images generated by the generator from the trained BEGAN (random noise as input) and CycleGAN. Then, we used these 10,000 images as training sets to train the extractor with a large number of random noise vectors. In the training procedure, we set the mini-batch to 100; Adam optimization is used in training, and the learning rate is set to 0.0002. We train the extractor for 200,000 steps. The loss function value of extractor was shown in Fig.10.

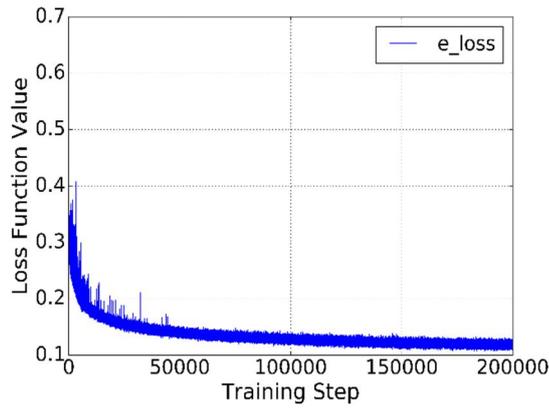

**Fig. 10** The loss function value of extractor in training

From the experimental results, it can be seen that the extractor converges rapidly and will eventually converge to a small value, which means that the output of the extractor is very close to the input noise vector. This means that we can reverse map the output of the extractor back to the secret message using Table 1.

*4.2 Experiments for GRDH implementation*

First, the secret information bit string (generated using random.randint in NumPy) is divided into several segments (3 bit/segment, i.e. $k$=3), and then the binary bit form segment is converted into a random noise form of the range [-1, 1] by Formula (4). Image generation, image transformation and image restoration were performed using the trained CycleGAN model and BEGAN model. Among them, CycleGAN model training image set is Man image set and Woman image set; the training frequency is 600,000 steps; BEGAN model training image set is CelebA image library; the training frequency is 400,000 steps; the noise dimension is set to 100, so the amount of embedded data in the experiment was set to 300 bit (i.e. 100×3). The experimental results of image generation, image transformation and image restoration in some samples are shown in Fig. 11(a). Due to the limitations of image quality generated by BEGAN, the visual quality of gender conversion and image restoration is relatively low. Fig. 11(b) shows some experimental results when the StyleGAN model is adopted in the image generation process.

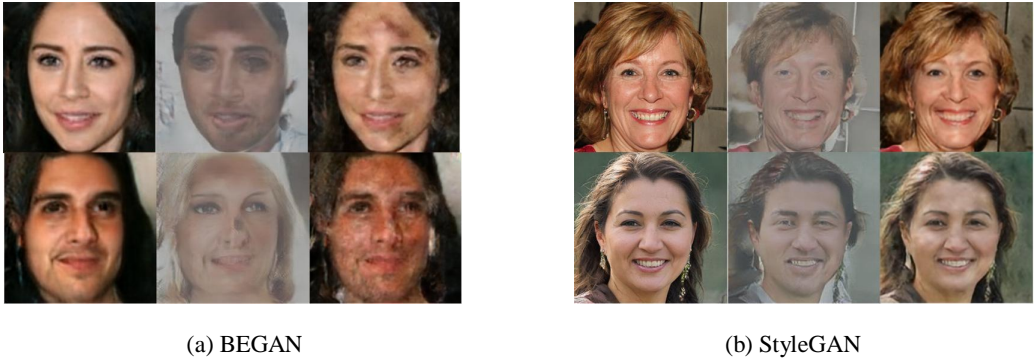

(a) BEGAN　　　　　　　　　　(b) StyleGAN

**Fig. 11** The results of Image generation, image conversion, and image restoration

In addition, we calculated the average value of peak signal-to-noise ratio (PSNR) for 1,000 recovered images from two GAN models separately, as shown in table 2.

**Table 2** The average values of PSNRs of recovered images

| GAN model | PSNR value (dB) |
|---|---|
| BEGAN | 22.665 |
| StyleGAN | 28.333 |

Although the average PSNR value of restored images is not high, the subjective visual effect is acceptable. Meanwhile, it is easy to see from the figure that the visual quality of gender conversion and image restoration is relatively high. Therefore, whether in the preparation phase or the implementation phase, the image visual quality is mainly affected by the type of the selected GAN model.

To test the extractor, we measured the accuracy of the extraction of secret information from 1,000 marked images. The average extraction accuracy of the extractor is 88.7%. In addition, to test the effect of parameters $k$ and $\delta$ on extractor recovery accuracy rate $R$, some further experiments are carried on. The effect of parameters $k$ and $\delta$ on recovery accuracy $R$ were shown in Table 3 and Table 4 respectively.

**Table 3** The effect of parameter $k$ on recovery accuracy $R$ when $\delta = 0.01$

| $k$ value | 1 | 2 | 3 | 4 | 5 |
|---|---|---|---|---|---|
| $R$ value | 0.951 | 0.937 | 0.891 | 0.765 | 0.657 |

**Table 4** The effect of parameter $\delta$ on recovery accuracy $R$ when $k = 3$

| $\delta$ value | 0.01 | 0.02 | 0.03 | 0.04 | 0.05 | 0.06 | 0.07 | 0.08 | 0.09 | 0.1 |
|---|---|---|---|---|---|---|---|---|---|---|
| $R$ value | 0.883 | 0.889 | 0.894 | 0.887 | 0.889 | 0.893 | 0.897 | 0.886 | 0.894 | 0.898 |

From the tables, the recovery accuracy $R$ significantly decreases with the rising of parameter $k$ and slightly increases with the rising of parameter $\delta$. To analyze the reason, it is just because that the smaller the parameter $k$ is, the better correction capability the algorithm will be. Although the extractor recovery accuracy is not perfect, this problem can be resolved by including error-correction codes in the input noise.

Because the generative steganography is to fully fit the data distribution in the natural image dataset through the training generator, this technique is very resistant to the machine learning based steganographic analyzer, which was mentioned in [18]. In other words, it is safer than traditional modification-based steganography. Besides, because of the data hiding principle, the embedding capacity is very limit now. In the future, with the progress of GAN model, the steganographic capacity of this scheme will gradually increase.

## 5 Conclusions

In this paper, a novel RDH scheme named GRDH based on GANs model is proposed. Firstly, we use the GAN model to train a powerful image generator to get realistic images. This image is imported into the CycleGAN model to obtain images with different semantic information. In order to achieve message embedding, then, we establish a mapping relationship between noise and messages. The extraction of the message is achieved by training an extractor to recover noise from the final dense image. Experimental results have demonstrated the effectiveness of the proposed method. Although the 100 percent reversibility cannot be achieved due to the existing performance of CycleGAN, the proposed method can achieve the first RDH scheme without cover modification. However, compared with the traditional methods, the embedding capacity of the proposed method is very limit but the security is higher. In future work, we will try to experiment with some new generative models to improve the quality of restored images and to improve the steganographic capacity of the method.

*Data Availability*

The data used to support the findings of this study are available from the corresponding author upon request.

*Conflicts of Interest*

The authors declare that there is no conflict of interest regarding the publication of this paper.

*Acknowledgments*

This work is partially supported by National Natural Science Foundation of China (No. 61379152，61403417 and 61402530), Shaanxi Provincial Natural Science Foundation (2014JQ8301).